\documentclass{emulateapj}

\def\1705{4U~1705--44}

\begin{document}

\title{A broad iron line in the Chandra/HETG spectrum of \1705}

\author{T. Di Salvo\altaffilmark{1}, R. Iaria\altaffilmark{1}, 
        M. M\'endez\altaffilmark{2}, L. Burderi\altaffilmark{3}, 
        G. Lavagetto\altaffilmark{1}, N. R. Robba\altaffilmark{1}, 
        L. Stella\altaffilmark{3}, M. van der Klis\altaffilmark{4}}

\altaffiltext{1}{Dipartimento di Scienze Fisiche ed Astronomiche, 
	      Universit\`a di Palermo, via Archirafi 36 - 90123 Palermo, 
              Italy; email: disalvo@gifco.fisica.unipa.it}
\altaffiltext{2}{National Institute for Space Research, Sorbonnelaan 2, 
              3584 CA Utrecht, the Netherlands}
\altaffiltext{3}{Osservatorio Astronomico di Roma, via Frascati 33, 
              00040 Monteporzio Catone (Roma), Italy} 
\altaffiltext{4}{Astronomical Institute ``Anton Pannekoek", University of 
	      Amsterdam and Center for High-Energy Astrophysics,
	      Kruislaan 403, NL 1098 SJ Amsterdam, the Netherlands}

\begin{abstract}

We present the results of a Chandra 30 ks observation of the low mass 
X-ray binary and atoll source \1705. Here we concentrate on the study of 
discrete features in the energy spectrum at energies below $\sim 3$ keV, 
as well as on the iron K$\alpha$ line, using the HETG spectrometer 
on board of the Chandra satellite. Below 3 keV, three narrow emission lines 
are found at $1.47$, $2.0$, and $2.6$ keV. The $1.47$ and $2.6$ keV are 
probably identified with Ly-$\alpha$ emission from Mg XII 
and S XVI, respectively. 
The identification of the feature at $\sim 2.0$ keV is uncertain 
due to the presence of an instrumental feature at the same energy. 
The iron K$\alpha$ line at $\sim 6.5$ keV is found to be intrinsically 
broad ($FWHM \sim 1.2$ keV); its width can be explained by reflection from 
a cold accretion disk extending down to $\sim 15$ km from the neutron star 
center or by Compton broadening in the external parts of a hot ($\sim 2$ keV) 
Comptonizing corona. We finally report here precise X-ray coordinates of the 
source.


\end{abstract}

\maketitle

\section{Introduction}


Broad emission  lines (FWHM  up to  $\sim 1$ keV)  at energies  in the
range 6.4 -- 6.7 keV are often observed in the spectra of Low Mass 
X-ray Binaries (hereafter LMXBs), both in systems containing black hole
candidates (see e.g.\ Miller et al.\ 2002) and in systems hosting an old 
accreting neutron star (see e.g.\ Barret 2001 for a review). 
These lines are identified with  the K$\alpha$ radiative transitions of
iron at different  ionization stages. Sometimes
an iron  absorption edge at energies  $\sim 7-8$ keV  has been detected.
These features are powerful tools to investigate the structure
of the accretion flow  close to  the central  source; in particular,
important information can be obtained from detailed
spectroscopy of the  iron K$\alpha$ emission line and absorption edge,
since  these  are determined by the ionization stage, geometry, and 
velocity field of the reprocessing  plasma.

To explain the large width of these  lines it has been proposed that  they
originate from emission reprocessed by the accretion disc surface
illuminated by the  primary  Comptonized spectrum (Fabian et al. 1989).
In  this  model, the combination of relativistic Doppler effects arising 
from the high orbital velocities  and gravitational effects  due to  the 
strong field in  the vicinity of the neutron star  smears 
the  reflected features. Therefore the  line will have a characteristically
broad profile, the detailed shape of which depends on the inclination
and  on  how deep  the  accretion disk  extends  into  the neutron star
potential (e.g.\ Fabian et al. 1989; Stella 1990).

An alternative location of the line emitting region is the inner part
of the so-called Accretion Disk Corona (ADC), probably formed by 
evaporation of the outer layers of the disk illuminated by the emission 
of the central object (e.g.\ White \& Holt 1982). 
In this case the width of the line is explained by
thermal Comptonization of the line photons in the ADC. This produces
a genuinely broad Gaussian distribution of line photons, with
$\sigma \geq E_{\it Fe} (k T_e/m_e c^2)^{1/2}$,
where $E_{\it Fe}$ is the centroid energy of the iron line and $k T_e$ is
the electron temperature in the ADC (see Kallman \& White 1989; 
Brandt \& Matt 1994 for more detailed calculations). This mechanism can
explain the width of the iron line for temperatures of the emitting
region of few keV.

The presence of several unresolved components, which can eventually be
resolved by the high resolution X-ray instruments on board Chandra and
XMM-Newton, can also contribute to broaden the line.


\1705 is a very interesting source of the atoll class (see Hasinger \& 
van der Klis 1989) which also shows type-I X-ray bursts. 
As in other similar sources, the energy spectrum of \1705 can be
described as the sum of a (dominating) Comptonized component, 
a blackbody, and an emission line at $\sim 6.4$ keV.
We selected this source for a Chandra observation because a broad
(1.1 keV FWHM) iron emission line at 6.5 keV has been previously reported
(White et al. 1986; Barret \& Olive 2002).
One of the goals of our Chandra observation was to study
the iron line profile to discriminate among the various models that
have been proposed to explain the large line width. The Chandra/HETGS
observation demonstrates that the iron line is intrinsically broad
(1.2 keV FWHM); possible broadening mechanism are discussed.

\section{Observations}

\1705 was observed using the High-Energy Transmission Grating Spectrometer
(HETGS) on board of Chandra starting on 2001 July 1. Part of the observation, 
for a total integration time of 24.4 ks, was performed in Timed Graded mode,
while a short part of the observation, $\sim 5$ ks, was performed in 
Continuous Clocking mode.
The HETGS consists of two types of transmission gratings, the Medium-Energy
Grating (MEG) and the High-Energy Grating (HEG). The HETGS affords high-resolution
spectroscopy from 1.2 to 31 \AA\ (0.4-10 keV) with a peak spectral resolution
of $\lambda / \Delta\lambda \sim 1000$ at 12 \AA\ for HEG first order.
The dispersed spectra were recorded with an array of six charged coupled
devices (CCDs) that are part of the Advanced CCD Imaging Spectrometer
(ACIS-S; Garmire et al. 2003; see http://asc.harvard.edu/cdo/about$\_$chandra 
for more details).
The current relative accuracy of the overall wavelength calibration is of 
order of 0.05\%, leading to a worst-case uncertainty of 0.004 \AA\ in the 
first-order MEG, and 0.006 \AA\ in the first-order HEG. We processed the 
event lists using available software (FTOOLS and the CIAO v.3.1 packages). 
We computed aspect-corrected exposure maps for each spectrum, allowing us to 
correct for effects from the effective area of the CCD spectrometer.

For the part of the observation performed in Timed Graded mode, the 
brightness of the source required additional efforts to mitigate 
``photon pile-up" effects. We applied a 400-raw ``subarray" (with the 
first raw = 1) during the observation that reduced the CCD frame time to 
1.4 s. The zeroth-order image is affected by heavy pile-up; the event rate
is so high that two or more events are detected in the CCD during the
1.4-s frame exposure.
Pile-up distorts the count spectrum because detected events overlap and
their deposited charges are collected into single, apparently more energetic,
events. Moreover, many events ($\sim 90\%$) are lost as the grades of the
piled-up events overlap those of highly energetic background particles,
and are thus rejected by the on-board software.
We therefore will ignore the zeroth-order events in subsequent analysis.
On the other hand, the grating spectra are not, or only moderately
(less than 10\%), affected by pile-up.  In our analysis, we utilize the
HEG first-order spectrum, which is less affected by pile-up (less than 6\%
at maximum, less that 3\% below 3 keV and above 6 keV) with respect to 
the MEG first-order spectrum.

To determine the zero-th point position in the image as precisely as possible
we calculated the mean crossing point of the zero-th order readout trace
and the tracks of dispersed HEG and MEG arms. This results in the following
source coordinates: RA = 17$^{\rm h}$ 08$^{\rm m}$ 54$^{\rm s}$.47,
DEC = $-44^\circ$ 06$'$ 07$''$.35 (J2000, uncertainty 0.5$''$). Note that 
this position is significantly different ($\sim 0.15$ arcmin) from the 
coordinates previously reported for this source (see Liu, van Paradijs, \& 
van den Heuvel 2001, and references therein).

The data collected in Continuous Clocking mode do not suffer from photon 
pile-up in the first order dispersed spectra. However, due to the short
exposure time of this observation, the statistics are quite low and we only 
used these data to check the results obtained with the Timed Graded 
observation.  Indeed the spectra acquired in Continuous Clocking mode are 
in good agreement with the HEG spectra in Timed Graded (both in the shape of 
the continuum emission and in the parameters of the iron line at 6.5 keV, 
see below). We therefore conclude that the effects of pile-up in the HEG 
spectra of \1705 acquired in Timed Graded mode are negligible.

\section{Spectral Analysis}

We selected the first-order spectra from the HEG. Data were
extracted from regions around the grating arms; to avoid overlapping
between HEG and MEG data we used a region size of 26 pixels for the HEG
along the cross-dispersion direction.  
The background spectra were computed, as usual, by extracting data 
above and below the dispersed flux. The
contribution from the background is $\sim 0.3\%$ of the total count rate.
We used the standard CIAO tools to create detector response files
(see Davis 2001) for the HEG $+1$ and $-1$ order (background-subtracted)
spectra, which we fit simultaneously using the XSPEC v.11.2 data
analysis package (Arnaud 1996). The HEG spectra were rebinned to $0.005 \AA$.

Two large energy gaps at those wavelength corresponding to the junction 
between two CCDs are present in the HEG first order spectra, at 3--3.4
keV and 5.5--6.5 keV for the +1 and --1 order, respectively. In these 
intervals the effective area is much smaller and is known with less 
accuracy (see Proposers' Observatory Guide, POG, at 
http://cxc.harvard.edu/helpdesk). 
Unfortunately in the interval between 6 and 6.6 keV, the HEG +1 and $-1$ 
order spectra give slightly different residuals with respect to the same 
continuum model. 
In the present work we therefore exclude the interval mentioned above 
corresponding to the junction between two CCDs and in which the two orders 
give different results. 
The energy ranges used in the following spectral analysis are: 1.3--10 keV 
for the $-1$ order, and 1.3--6 keV and 6.6--10 keV for the +1 order,
respectively.

We fit the HEG first order spectra of \1705 to a continuum model. The best
fit model consists of the Comptonization model {\tt comptt}
(Titarchuk 1994), modified by absorption from neutral matter, parametrized
by the equivalent hydrogen column $N_H$, which gives a
$\chi^2_{\rm red} (d.o.f.)$ of $1.05 (3265)$. This model also includes
an overabundance of Si by a factor $\sim 2$ with respect to Solar
abundances to fit a highly significant absorption edge at
$\sim 1.84$ keV (the addition of this parameter reduces the $\chi^2$ by 
$\Delta \chi^2 \simeq 61$ at the expense of 1 degree of freedom). 
Note that we cannot exclude that this feature may be due to the presence of 
Si in the CCDs, and therefore this overabundance is not discussed further. 
Finally, in all the fits we include an 
instrumental feature at $2.06$ keV (usually present in the HETG spectra 
of bright sources, see Miller et al. 2002) which is fitted by an inverse 
edge (with $\tau \sim -0.1$).

A soft blackbody component has often been reported in the X-ray spectra of
this kind of sources and has been detected in the spectra of \1705 as well. 
In \1705, in spectral states similar to the one found during our Chandra 
observation, this component has a temperature of $\sim 1.9$ keV and can 
contribute up to 20\% of the total source flux (Barret \& Olive 2002). 
We therefore added a blackbody component to the {\tt comptt} continuum 
model; this component has a  temperature of $\sim 1.95$ keV and contributes
up to 50\% of the total X-ray flux. However, the decrease of the $\chi^2$
for the addition of this component ($\Delta \chi^2 = 5$ for the addition 
of two parameters) is not significant, and, for sake of simplicity, we 
preferred not to include this component in our continuum model.

\begin{figure}[h!]
\plotone{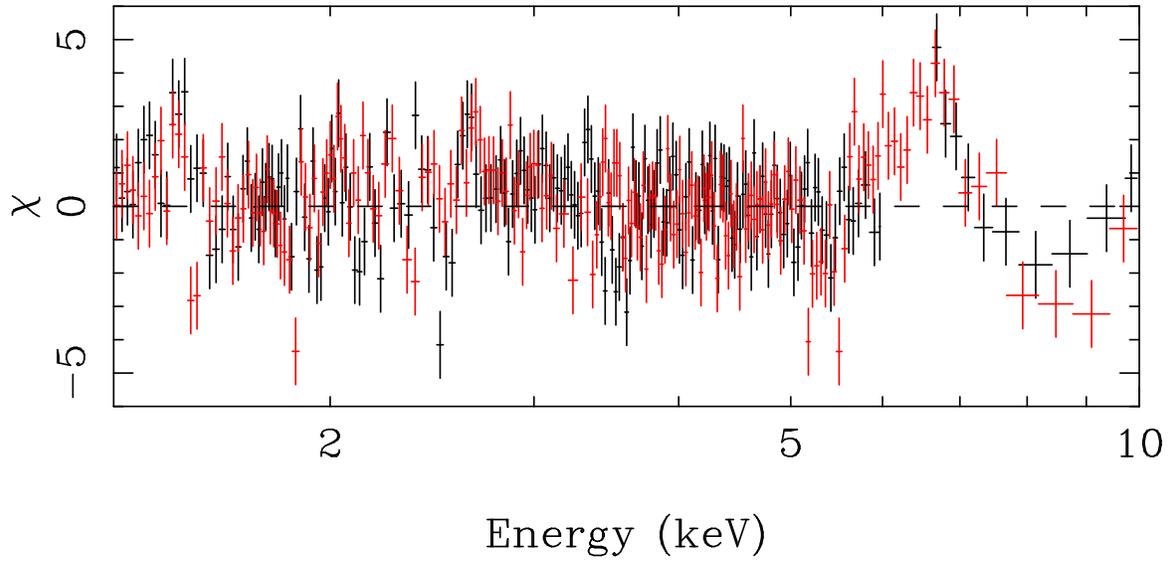}
\caption{Residuals in units of $\sigma$ with respect to the simple
Comptonization continuum when the iron line and other low energy Gaussians 
are not included in the model.}
\label{res}
\end{figure}

Residuals in units of $\sigma$ with respect to the continuum model
described above are shown in Figure~\ref{res}; several discrete features are
still clearly visible in the residuals with respect to this continuum
model, at $\sim 1.5$, $2.0$, $2.6$ keV, and, particularly, in the 6--7 keV 
range, where the K$\alpha$ iron emission line is expected. 
From these residuals it is apparent that the iron line is intrinsically broad 
and shows a complex profile.

\begin{figure}[h!]
\plotone{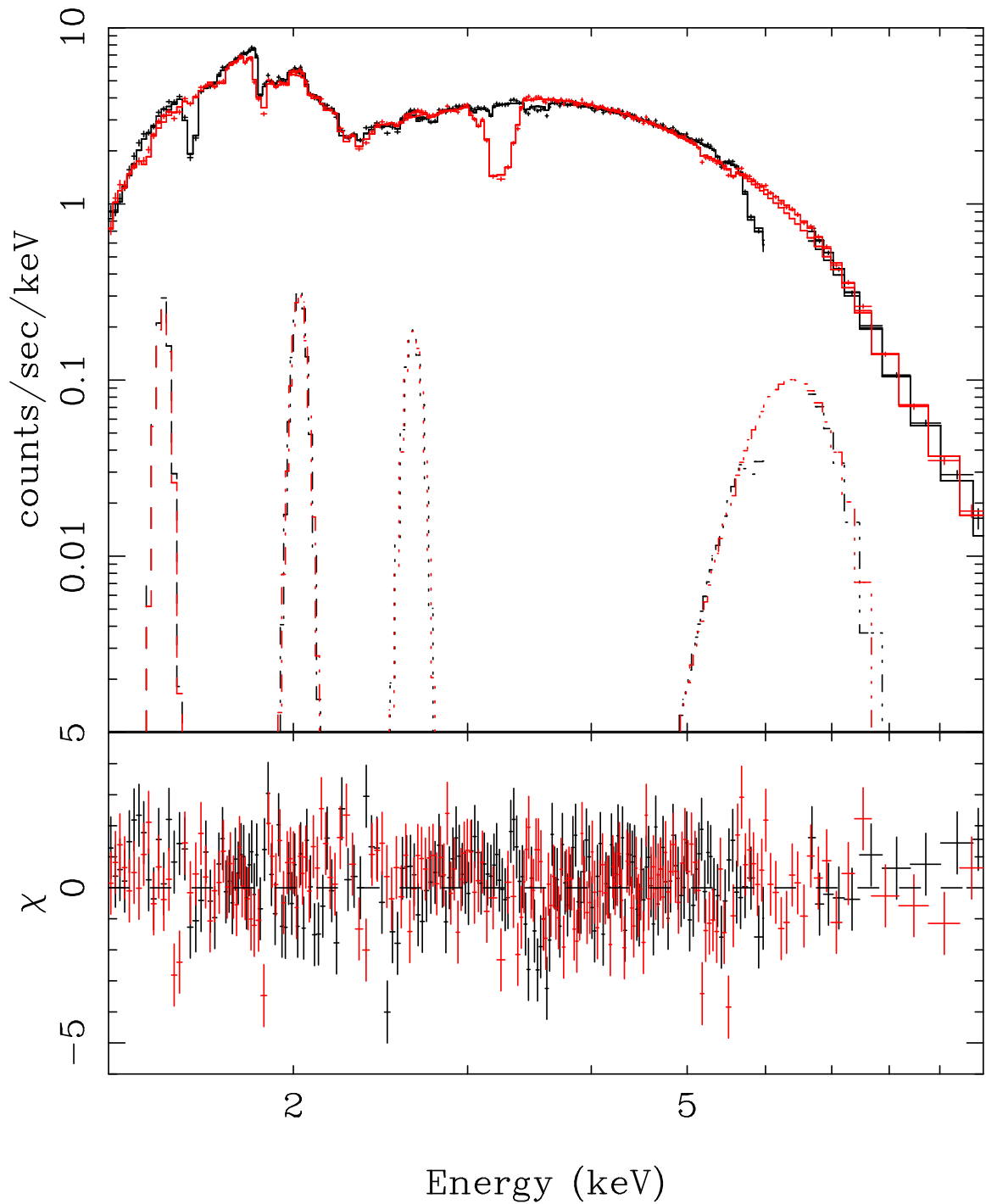}
\caption{Top panel: HEG first order spectra of \1705 together with the 
best-fit model (see Table 1). The discrete features included in the
best-fit model and described by Gaussians are also shown. Note that,
for seek of clarity, we have rebinned the data in the plot with
respect to the energy resolution used for the spectral analysis. 
Bottom panel: Residuals in units of $\sigma$ with respect to the best fit 
model.}
\label{spect}
\end{figure}

The addition of a broad ($\sigma \sim 0.5$ keV) Gaussian line centered
at 6.5 keV proves necessary, giving $\Delta \chi^2 = 167$ for the 
addition of three parameters. We also added three narrow emission lines
to fit the other low energy residuals mentioned above.
The addition of Gaussian emission lines at $\sim 1.5$, $2.0$, $2.6$ keV
gives a reduction of the $\chi^2$ by 27, 29, and 32 units, respectively,
for the addition of three parameters. The errors in the normalizations of these 
features give a detection at about $3 \sigma$ confidence level. This is not a 
highly significant detection, and needs a confirmation with future 
observations. However, the fact that the energies of these features are 
close to the energies of Ly$\alpha$ transitions of H-like ionization stages
of the most abundant ions emitting in the observed range (that are Mg XII,
Si XIV, and S XVI, respectively) adds further confidence that these lines may 
be real.
Data and residuals in units of $\sigma$ with respect to this best-fit model 
are shown in Figure~\ref{spect}.  The best fit model is reported in Table 1 
together with the discrete features mentioned above as well as
the identification of the line, when possible.

The spectral analysis of the first order HEG spectra of \1705 from the 
part ($\sim 5$ ks) of the Chandra observation performed in Continuous 
Clocking mode confirms the best fit model found for the Timed Graded 
spectra. In particular, the best fit continuum model is again given by
the {\tt comptt} model, which gives a $\chi^2/dof = 867/1097$. The addition
of a broad Gaussian emission line at $6.6 \pm 0.1$ keV ($\sigma = 0.4 
\pm 0.1$ keV, EW = 138 eV) improves the fit giving a  
$\chi^2/dof = 831/1094$.


The residuals in the K-shell iron line range are quite complex (see 
Fig.~\ref{res}),
showing a broad feature centered at $\sim 6.5$ keV. We therefore try
to fit this feature with the line profile expected from a thin Keplerian
accretion disk. Substituting the Gaussian line with the {\tt diskline} 
model (Fabian et al.\ 1989), we obtain an equivalently good fit, 
$\chi^2(d.o.f.) = 3164/3252$ using the {\tt diskline} model.
The line best fit parameters for the {\tt diskline} model are given in Table 2.

\section{Discussion}

We have analysed a Chandra 30 ks observation of the X-ray burster and 
atoll source \1705.  The position of the zero-th order image of the source
provides improved X-ray coordinates for \1705 (RA = 17$^{\rm h}$ 
08$^{\rm m}$ 54$^{\rm s}$.47, DEC = -44$^\circ$ 06$'$ 07$''$.35), which 
significantly differ (by about 0.15 arcmin) from the coordinates 
previously reported for this source.

We performed a spectral analysis of the HEG first order spectra of \1705.
The continuum emission is well fitted by the Comptonization model 
{\tt comptt}, with an equivalent hydrogen column of $\sim 1.4 \times 
10^{22}$ cm$^{-2}$, and an overabundance of Si by a factor $\sim 2$ with
respect to Solar abundance (which might be of instrumental origin).
The inferred unabsorbed flux of the source in the 0.1--10 keV range 
is $\sim 1.0 \times 10^{-8}$ ergs cm$^{-2}$ s$^{-1}$, corresponding to
a luminosity of $3.3 \times 10^{37}$ ergs s$^{-1}$ assuming a distance 
to the source of 7.4 kpc (Haberl \& Titarchuk 1995).
The Comptonization continuum is quite soft, with an electron temperature
of $k T_e \sim 2.3$ keV and an optical depth of $\tau \sim 18$ for a 
spherical geometry. 

We have detected a broad emission feature at $6.4-6.5$ keV, which we interpret 
as K-shell fluorescent emission of lowly ionized iron.
The high energy resolution of the HEG shows that the line is intrinsically
broad ($FWHM \sim 1.2$ keV), in agreement with previous measures (see e.g.\
Barret \& Olive 2002). Therefore the most probable origin of this line
is an accretion disk (in this case the large width of the line would
be due to Doppler and relativistic smearing effects) or a hot
corona (in this case the large width of the line would be due to
Compton broadening).

In the case the line is produced by reflection in an accretion disk we
estimate that the required inner radius of the disk is $\sim 7\; R_g$
or $\sim 15$ km for a $1.4\; M_\odot$ neutron star. Note that the quite
small inner radius of the disk inferred from this model is in agreement 
with the quite soft X-ray spectrum of \1705 during the Chandra observation,
which would probably place the source in the banana state of its X-ray
color-color diagram.
In this model, the inclination of the disk with respect to the line of 
sight is constrained in the range $55^\circ - 84^\circ$.
Alternatively, Comptonization in the corona could explain the large width
of the line.  Detailed calculations give
$\sigma_{\rm Fe} = 0.019 E_{\it Fe} \tau_T (1 + 0.78 kT_e)$,
where $\tau_T$ is the Thompson optical depth and $kT_e$ is in keV
(Kallman \& White 1989, see also Brandt \& Matt 1994).
Assuming an average electron temperature of $k T_e = 2.3$ keV,
as derived from the fit of our data to the Comptonization model, we can
explain the width of the iron line for a Thomson optical depth of
$\tau_T \sim 1.4$. Therefore, it is possible that the line is produced
in the outer region of the Comptonizing corona, where the optical depth
might be lower (assuming that the temperature remains constant).
This is not unreasonable given that any contribution to the line produced
inside the Comptonizing region, where the optical depth can be as high as 10
(see Tab. 1), would be completely smeared by Comptonization.
Therefore, as expected, we only see that part of the line that is produced
in the outer Comptonizing region.

Unfortunately with these data we are not yet able to discriminate between 
the two possible origins of the iron line and its broadening proposed above. 
Higher statistics or detection of line variability would be needed for this.
Naturally, what would definitively discriminate between the two proposed 
models is the detection of a double peak in the iron line profile; this 
would exclude the Comptonization of line photons model and would indicate 
the relativistic/Doppler effects as the origin of the line broadening. 
This detection would be possible with higher statistics (large effective areas 
or long exposures). Hence, more XMM-Newton and Chandra observations (better 
if simultaneous) would be ideal to this aim. 
Snapshots observations at different intensity levels and/or spectral states
of the source would also be useful to study the variability of the line 
parameters with the position of the source in the X-ray color-color diagram 
and/or the frequency of the so-called kiloHertz quasi-periodic oscillations
(kHz QPOs). 
If the source position in the color-color diagram track is indeed determined 
by the mass accretion rate (see e.g.\ Hasinger \& van der Klis 1989) and the 
frequencies of the kHz QPOs are related to the Keplerian frequency at the inner 
edge of the accretion disk (as envisaged by most of the current models, 
see e.g.\ Miller, Lamb, \& Psaltis 1998; Stella \& Vietri 1999), then one 
would expect that the radius of the disk as determined from the line profile 
(fitted with a {\tt diskline} model) should change accordingly. These kind of 
studies could give other pieces of evidence in favour of one or the other model.

To fit discrete residuals at low energies with respect to the continuum 
model we added to the model several Gaussian lines. In Table 1 we report 
the most significant of these features, together with a possible 
identification of the line. The lines at $1.5$ and $2.6$ keV are identified 
with Ly$\alpha$ transitions from H-like Mg at 1.4726 keV (or, less probably, 
L-shell transitions from highly ionized iron, Fe XXII -- Fe XXIV) 
and S at 2.6227 keV, respectively. 
The identification of the line at $\sim 2.0$ keV is more uncertain due to the 
presence of a systematic feature at $\sim 2.07$ keV. Note, however, that 
this line is quite close to the Ly$\alpha$ transition of Si XIV at 
2.0061 keV. 
Finally, we note that there seems to be a correlation between the energy and 
the width of these lines, with the Gaussian $\sigma$ increasing with the 
centroid energy of the line.
If the line width is due to velocity dispersions, we have calculated that 
$\Delta E/ E_0 = v/c \sim 2.7 - 3.6 \%$ (where we have used for $\Delta E$
the FWHM), slightly increasing with the energy of the line, 
from Mg XII to S XVI, as expected if more ionized elements are produced 
closer to the central X-ray source.

\acknowledgements
This work was partially supported by the 
Ministero della Istruzione, della Universit\'a e della Ricerca (MIUR).

\clearpage

\clearpage

\clearpage

\begin{table}[h!]
\footnotesize
\caption{Results of the fitting of the \1705 HEG first order spectra in the
1.3--10~keV energy band.  }
\label{table:1}
\newcommand{\m}{\hphantom{$-$}}
\newcommand{\cc}[1]{\multicolumn{1}{c}{#1}}
\renewcommand{\tabcolsep}{1.5pc} 
\renewcommand{\arraystretch}{1.2} 
\begin{center}
\begin{tabular}{@{}ll}
\hline
Parameter               & Value \\
\hline
$N_{\rm H}$ $\rm (\times 10^{22}\;cm^{-2})$ & $1.42 \pm 0.06$ \\
Si / Si$_\odot$				    & $2.0 \pm 0.2$ \\
$k T_0$ (keV)                               & $0.50 \pm 0.02$ \\
$k T_{\rm e}$ (keV)                         & $2.29 \pm 0.09$  \\
$\tau$                                      & $17.7 \pm 0.7$  \\
Flux (1.3--10 keV, erg cm$^{-2}$ s$^{-1}$)  & $7.82 \times 10^{-9}$ \\
Final $\chi^2(d.o.f.)$			    & $3168 / 3255$ \\
\hline
$E_1$ (keV, ID: Mg XII Ly$\alpha$)          & $1.476 \pm 0.007$ \\
$\sigma_1$ (eV)                             & $17 \pm 6$ \\
I$_1$ ($10^{-3}$ cm$^{-2}$ s$^{-1}$)        & $2.3 \pm 0.9$  \\
EW$_1$ (eV)                                 & $ 4.28 $ \\

$E_2$ (keV, ID: Si XIV Ly$\alpha$ ?)        & $2.03 \pm 0.01$ \\
$\sigma_2$ (eV)                             & $28 \pm 10$ \\
I$_2$ ($10^{-3}$ cm$^{-2}$ s$^{-1}$)        & $1.9 \pm 0.7$  \\
EW$_2$ (eV)                                 & $ 4.17 $ \\

$E_3$ (keV, ID: S XVI Ly$\alpha$)           & $2.64 \pm 0.02$ \\
$\sigma_3$ (eV)                             & $40 \pm 14$ \\
I$_3$ ($10^{-3}$ cm$^{-2}$ s$^{-1}$)        & $2.3 \pm 0.8$  \\
EW$_3$ (eV)                                 & $  6.32 $ \\

$E_{\rm Fe}$ (keV)                          & $6.54 \pm 0.07$ \\
$\sigma_{\rm Fe}$ (keV)                     & $0.51 \pm 0.08$ \\
I$_{\rm Fe}$ ($10^{-2}$ cm$^{-2}$ s$^{-1}$) & $1.5 \pm 0.3$  \\
EW$_{\rm Fe}$ (eV)                          & $ 170 $ \\
\hline
\end{tabular}\\[2pt]
\end{center}
The model consists of a Comptonized spectrum modeled by {\tt comptt}, and 
four Gaussian emission lines.
$k T_0$ is the temperature of the seed photon (Wien) spectrum, $k T_e$ the
electron temperature and $\tau$ the optical depth in a spherical geometry.
For the discrete features, $I$ is the intensity of the line and $EW$ is 
the corresponding equivalent width.
Uncertainties are given at 90\% confidence level.
\end{table}

\clearpage

\begin{table}[htb]
\caption{Iron line parameters from the {\tt diskline} model.}
\label{table:2}
\newcommand{\m}{\hphantom{$-$}}
\newcommand{\cc}[1]{\multicolumn{1}{c}{#1}}
\renewcommand{\tabcolsep}{2pc} 
\renewcommand{\arraystretch}{1.2} 
\begin{center}
\begin{tabular}{@{}ll}
\hline
Parameter               & Value \\
\hline
Energy (keV)            & $6.40 \pm 0.04$ \\
R$_{\rm in}$ (R$_g$)    & $7^{+4}_{-1}$ $(< 11)$ \\
R$_{\rm out}$ (R$_g$)   & $410^{+230}_{-130}$ \\
Inclination (deg)       & $59^{+25}_{-4}$ \\
Index                   & $2.1 \pm 0.2$ \\
I ($10^{-2}$ cm$^{-2}$ s$^{-1}$) & $1.8 \pm 0.3$ \\
Final $\chi^2(d.o.f.)$  & $3164/3252$ \\
\hline
\end{tabular}\\[2pt]
\end{center}
The other best fit parameters are compatible with those reported in Table 1.
Index refers to the power-law dependence of emissivity which scales
as $r^{-\rm Index}$.
Uncertainties are 90\% confidence level for a single parameter of interest.
\end{table}

\end{document}